**Early steps in the formation of the interface between organic molecular semiconductors and metals: a computational approach**

*Francesco Mercuri\**

Istituto per lo Studio dei Materiali Nanostrutturati (ISMN), Consiglio Nazionale delle Ricerche (CNR)
Via P. Gobetti 101, 40129 Bologna, Italy
E-mail: francesco.mercuri@cnr.it

A computational approach for predictive simulations of the nanoscale morphology in the early steps of the formation of the interface between metals and organic molecular semiconductors is presented. Despite the relevance of the metal-molecule junction for the development of electronic applications, structural details at the interface are often difficult to assess. Our approach, based on the integration of density functional theory with methods for the simulation of growth dynamics, allows to unravel the structural details of the formation of gold aggregates onto molecular materials. Simulations are applied to investigations of the initial steps in the formation of gold clusters at the interface with prototypical *p*-type and *n*-type organic molecular semiconductors. Results show a striking correlation between the morphology of the metal-organic interface, the details of the molecular structure and the peculiar metal-molecule interaction, also highlighting the role of fabrication conditions.

The performances of devices based on organic molecular materials, such as organic light-emitting diodes (OLEDs) or organic thin-film transistors (OTFTs), are critically related to the intrinsic properties of materials and to interface phenomena.[1] In particular, the properties of the interface between organic semiconductor materials and electrodes, usually constituted by metallic materials, affect dramatically the performance of devices, impacting for example in charge injection rates.[2–4] The main features of the metal-organic interaction are related to the morphology of the interface, which depends on the specific fabrication and processing conditions,[5,6] and to the resulting electronic properties, in terms of energy level alignment, charge transfer, formation of dipoles and interface states, and overlap of the electronic states.[7,8] In addition to experimental work, several theoretical studies provided models of organic molecules interacting with metallic surfaces, addressing the structure and the electronic properties at the interface.[9–12] These investigations clarified the strong dependence of the electronic properties of the interface in terms of the specific metal-organic interactions. However, basic models of the formation of the metal-organic interface in the growth of metal layers onto organic materials are missing.

In this work, we report on computational investigations of the initial steps in the formation of the interface between an organic molecular layer and a metal. In particular, we address the mechanisms underlying the formation and growth of metallic layers onto organic molecules, generally achieved by vacuum sublimation processes, as occurring in the fabrication of metal contacts over organic thin-films. By reproducing the morphology of metal nanoaggregates in contact with organic molecules, our simulations highlight the occurrence of relevant structural and electronic phenomena at the interface. We target the interface between gold, an archetypal materials for electrodes, and state-of-the-art organic semiconductor molecules, α,ω-dihexyl-quaterthiophene (DH4T) and N,N-dioctyl-3,4,9,10-perylenedicarboximide (PTCDI-C8).[13–15] Molecules based on oligothiophenes are ubiquitously used as *p*-type semiconductor materials. In particular, in DH4T the presence of thiophene cores is known to enhance the propensity to interaction with gold atoms and clusters.[16,17] PTCDI-C8, a molecule based on the perylene core,[18,19] is commonly used as an *n*-type semiconductor.[20] Despite the limited chemical affinity between gold and the π-core, molecules based on the perylene moiety often exhibit excellent electron injection and transport properties in devices. We therefore investigate how gold metal clusters interact with organic molecules and how the early steps in the formation of metal clusters in contact with the molecule are related to the peculiar chemical properties of the molecule. We apply electronic structure calculations based on density functional theory (DFT) and semiempirical methods, coupled to computational techniques derived from kinetic Monte Carlo (kMC) methods to simulate the growth of metal clusters in contact with organic molecules as a function of the processing conditions. Our simulations suggest strikingly different interface morphologies for the two molecules under investigation. In particular, the occurrence of molecular moieties that exhibit a strong affinity towards gold atoms induce the localized growth of molecular clusters and a strong renormalization of the electronic structure at the interface, resulting in the manifestation of interface electronic states. In contrast, molecules with low affinity towards gold atoms lead to less localized interactions at the interface and to delocalized growth of gold clusters onto the organic layer.

In a first set of simulations, we investigated the energetics involved in the initial steps of formation of the gold-molecule interface, as for example occurring during the thermal evaporation of gold contacts on organic molecular materials. Target organic molecules were initially optimized by DFT. Then, a 3D mesh of equispaced points was generated around the molecule, as described in the Methods section. The total energy for a single gold atom in each of the grid points was computed at both the DFT and semiempirical level. The volume map representing the interpolated gold-molecule interaction energies computed at the DFT level on the grid points is shown in **Figure 1**.

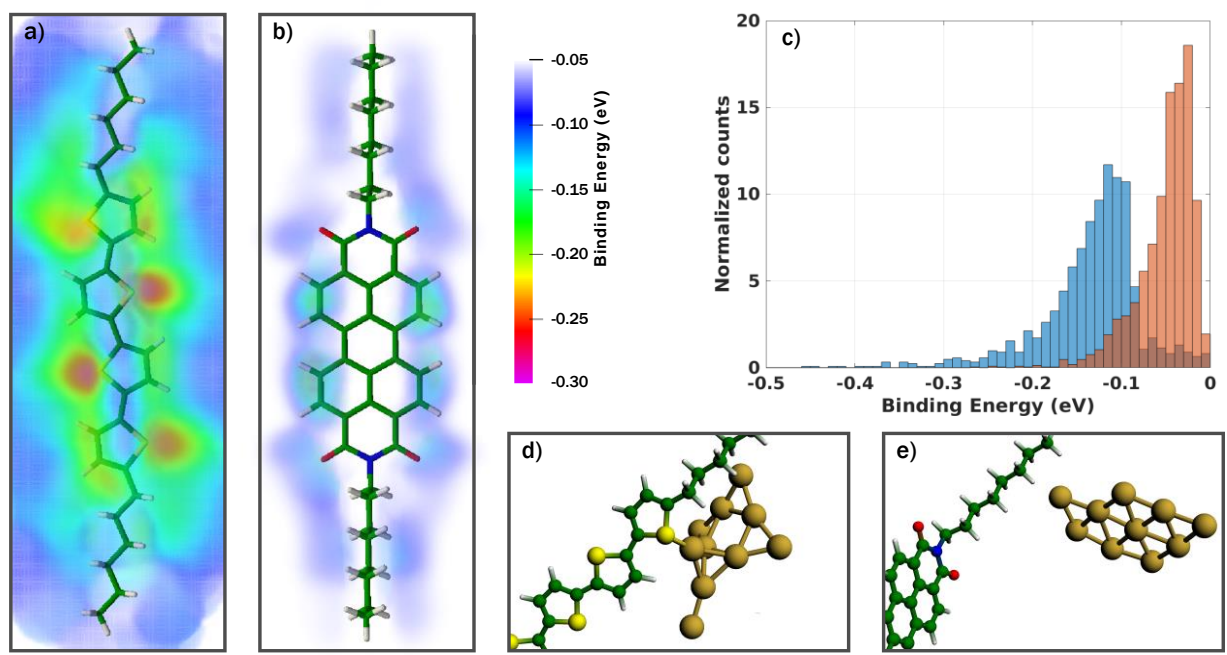

**Figure 1.** Volume rendering of the gold-molecule interaction energies (see scale bar) computed at the DFT level for a) DH4T and b) PTCDI-C8. c) Histogram of the binding energies computed at the mesh grid points for DH4T (blue bars) and PTCDI-C8 (red bars). Zoom of the minimum energy configurations for the d) $Au_{10}$DH4T and e) $Au_{10}$PTCDI-C8 systems.

The spatial distribution of the interaction energies suggests the occurrence of strongly localized regions of low-energy gold-molecule configurations in DH4T (see Figure 1a). As expected, these regions occur in the proximity of sulfur atoms and correspond roughly to high density areas of the LUMO of the thiophene rings. For the unrelaxed configurations, the resulting minimum binding energy of the Au-DH4T configurations extracted from the grid mesh, at the DFT level, is about -0.46 eV. Conversely, the spatial distribution of interaction energies for a gold atom interacting with a PTCDI-C8 molecule does not show evidence of a strongly preferential interaction region (see Figure 1b). In this case, the minimum interaction energy within the selected grid points lowers to -0.27 eV and occurs in correspondence of the *ortho-* carbon atoms of the perylene moiety. The overall distribution of binding energies around the molecule is shifted towards stronger interactions for DH4T with respect to PTCDI-C8, as highlighted in Figure 1c. DFT relaxation of the Au position for minimum energy configurations led to binding energies of -0.50 eV for DH4T and is essentially unchanged (- 0.28 eV) for PTCDI-C8, respectively. The computed binding energy between DH4T and a gold atom is in line with earlier theoretical work[21] and with values extracted from electron spectroscopy for the interaction between thiophene on gold surfaces (-0.48 eV) in configurations dominated by the sulfur-gold interaction.[22] Interestingly, calculations also provide indication of a sizeable charge transfer from the molecule to the gold atom for DH4T (about $0.21e$), as recently observed in similar systems,[23,24] and that is negligible ($0.02e$) in PTCDI-C8. The fractional net charge transfer between DH4T and gold is assisted by the electropositive (H and C) atoms of the molecule[25] and is a fingerprint of a strong hybridization between an organic molecule and a metal, as observed in other systems.[26] Although the interaction between gold atoms and the DH4T molecule alters significantly the electronic density distribution at the interface, the overall structure of the organic molecule is essentially unchanged, as observed experimentally.[22,27,28] The initial steps in the formation of the gold-molecule interface are therefore expected to follow different mechanisms in these two prototypical cases. In DH4T, the occurrence of thiophene rings induces a strong localization of the gold atoms in the proximity of sulfur atoms, with a significant renormalization of the molecular electronic density.[29,30] In PTCDI-C8, the gold-molecule interaction is weaker and spatially more uniformly distributed around the molecule. The gold-molecule interaction energies in the selected grid points were also calculated at the semiempirical (PM6) level (see **Figure S1** in the Supporting Information). The overall topology of the distribution of the PM6 interaction energies around the molecules reproduces the main features of the DFT calculations. In particular, the regions of strong gold-molecule interaction localized in correspondence of the $p$ orbitals of the sulfur atom are correctly described also at the PM6 level. Therefore, the preferred regions for the interaction between gold atoms and the molecule can be initially screened by PM6 calculations, and subsequently refined by DFT optimizations.

The simulation protocol described above was further extended by adding, one by one, other gold atoms to the molecule. At each iteration, the energetics of the gold-molecule system was computed at each point of the 3D mesh grid around the molecule at the PM6 level and the minimum energy configurations evaluated. For these latter, the position of the incoming gold atom was refined by constrained DFT optimization. The procedure was repeated adding up to ten gold atoms. This approach is expected to mimic the structure of minimum energy configurations occurring in the early steps of the formation of the gold-molecule interface. The relaxed configurations obtained after adding ten gold atoms to DH4T and PTCDI-C8 molecules are shown in **Figure 1d-e**. In DH4T, the gradual introduction of gold atoms to the system leads to the growth of a gold cluster anchored to the sulfur atom of one of the thiophene rings through an Au-S bond. The structure of the gold cluster corresponds to that expected for $Au_8$,[31,32] with an extra in-plane Au atom and an out-of-plane Au atom bound to the sulfur atom. The formation of a gold cluster is observed also in the case of PTCDI-C8. Here, the initial minimum energy configuration corresponds to a gold atom in the proximity of the alkyl chain of the PTCDI-C8 molecule. Subsequently, incoming gold atoms find a favorable configuration near the first gold atom and tend to form a cluster. The structure of the cluster is essentially planar and very close to that expected for an isolated $Au_{10}$ system.[33,34] However, as shown in Figure 1b, the energy landscape for a single Au atom interacting with a PTCDI-C8 molecule is rather shallow, and different regions for the nucleation of gold clusters around the molecule can in principle be expected. The average nearest-neighbor Au-Au distance within the cluster in contact with the molecules (2.78 Å and 2.75 Å for DH4T and PTCDI-C8, respectively) is in excellent agreement with the bond length observed in small Au clusters.[35]

The role of kinetic effects in the initial steps of formation of the Au-molecule interface was assessed by applying a method derived from kMC, as described in the Methods section. The inclusion of kinetic effects allows to model the processes related to the formation of the gold-molecule interface under thermodynamic and kinetic control, respectively, thus mimicking different experimental conditions. The initial steps in the formation of gold aggregates at the interface with DH4T and PTCDI-C8 were simulated at two different temperatures (300K and 1000K), which have been chosen in order to emphasize configurations corresponding to thermodynamic and kinetic conditions, respectively. In each aggregation simulation five Au atoms were consecutively added to the molecule, and each simulation was repeated ten times, with different starting random conditions. The equilibrium position of gold atoms was averaged over the whole set of simulations, obtaining a distribution of the likely regions for favorable gold-molecule interaction. Despite the limited statistical sampling, this approach is able to capture the significantly different behavior of the two target molecules considered in terms of formation of the gold-molecule interface in thermodynamic or kinetic control. The averaged distribution of the gold atoms in the early steps of the formation of the interface with DH4T and PTCDI-C8 in kinetic conditions is shown in **Figure 2a-b**, the corresponding distributions obtained in thermodynamic conditions are reported in **Figure S2** in the Supporting Information.

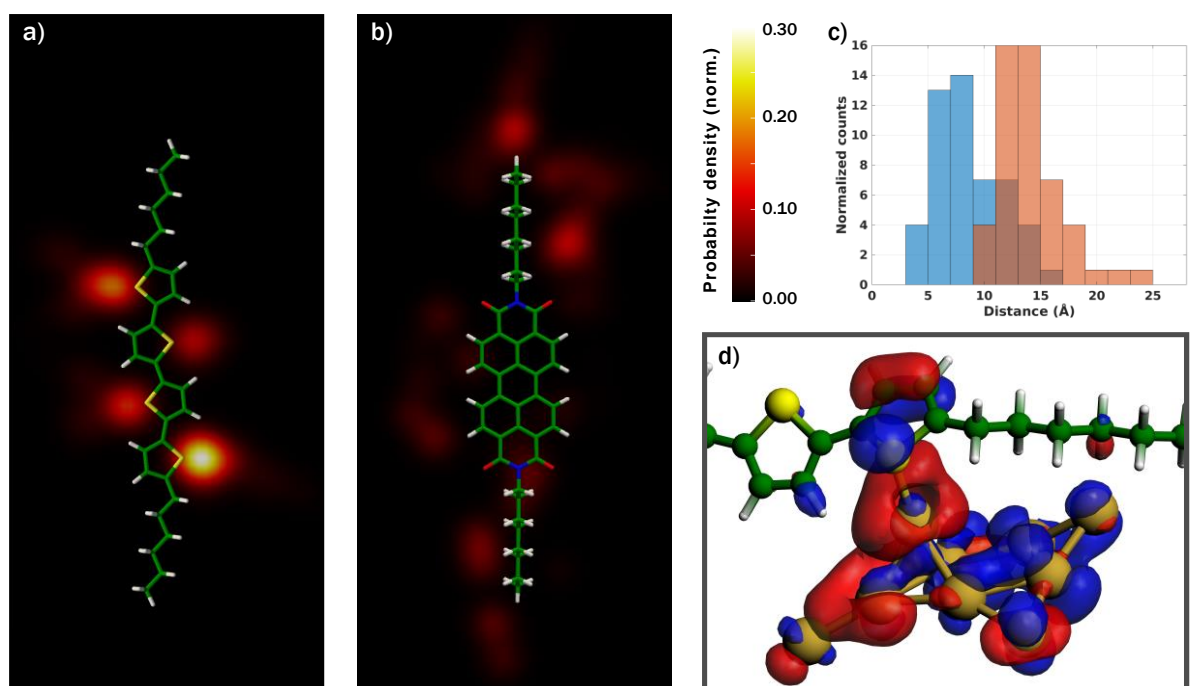

**Figure 2.** Projection on a plane parallel to the molecular π-core of the distribution of positions of gold atoms in the formation of the interface with a) DH4T and b) PTCDI-C8, averaged

over the whole ensemble of simulations, in kinetic conditions. c) Histogram of the distance of gold atoms with respect to their center of mass, in kinetic conditions, in contact with DH4T (blue bars) and PTCDI-C8 (red bars). d) Topology of a bonding molecular orbital, at an energy of -6.9 eV, for an $Au_{10}$ gold cluster interacting with a DH4T molecule in the minimum energy configuration.

In DH4T (Figure 2a and Figure S2a in the Supporting Information), both in thermodynamic and kinetic conditions, all simulations converge to similar configurations, which correspond to a gold cluster in contact with one of the thiophene rings through a gold-sulfur bond. The formation of the gold-molecule interface is therefore dominated by a deep free energy basin, related to gold-thiophene interactions, which leads to a strongly localized region for the nucleation and subsequent growth of a gold cluster. In the case of PTCDI-C8 (Figure 2b and Figure S2b in the Supporting Information), the weaker gold-molecule interaction leads to more delocalized regions for the growth of gold clusters in the early steps of the formation of the interface. Notably, growth simulations performed in thermodynamically- or kinetically-control, respectively, exhibit qualitatively similar features for the two molecules considered, suggesting in this case a minor role of the growth conditions in the formation of the gold-molecule interface. The distributions of the distances of gold atoms with respect to their center of mass, computed for the whole set of simulations (Figure 2c), also highlight a greater propensity to formation of compact aggregates for gold clusters interacting with DH4T compared to PTCDI-C8. The nature of the interaction between gold atoms and clusters and organic molecules is also reflected in the distribution of the resulting electronic states. The weaker interaction between PTCDI-C8 and gold leads essentially to a polarization of the molecular electronic states upon formation of the metal-organic interface (see **Figure S3** in the Supporting Information). The formation of a stronger gold-molecule chemical bond in DH4T (see Figure 2d), conversely, leads to a strong renormalization of the molecular electronic levels at the interface (see Figure S3 in the Supporting Information). It is worth noting that, despite the well-known quantitative inaccuracies in the description of energy level alignments for weakly interacting systems,[36,37] DFT methods are expected to reproduce accurately the formation of chemical bonds.

In summary, we presented a computational approach to simulate the early steps of formation of gold-molecule interfaces, applied to the case of two prototypical organic semiconductor materials for electronics. The morphology of the gold-molecule interface is indeed crucial in several technological processes related to the use of organic molecular materials in applications, such as charge injection from electrodes into a molecular thin-film. Simulations indicate a strikingly different behavior for the two molecules under investigation. In DH4T, the occurrence of thiophene rings induces the nucleation of gold clusters in strongly localized regions of the molecule, through strong gold-sulfur interactions. In DH4T, the growth of the gold-molecule interface proceeds by formation of gold nanoclusters in the proximity of the initial nucleation regions. Moreover, the hybridization between the orbitals of DH4T and the metal states leads to formation of interface states and to fractional charge transfer from the molecule to the gold atoms. Conversely, gold atoms tend to interact loosely with PTCDI-C8 molecules, and the formation of the gold-molecule interface in PTCDI-C8 reflects the growth of isolated gold clusters around the molecule. Our study clarifies the role of specific functionalities in organic semiconductor materials interacting with metallic layers during the deposition of electrodes over the organic layers. These findings highlight the need for a detailed description of interface phenomena, at the molecular level and at the nanoscale, to engineer metal-organic interfaces, materials, and processing and fabrication conditions.

**Methods Section**

*Computational Details*: Electronic structure calculations on selected configurations and geometry optimizations were performed at the gradient-corrected spin-unrestricted DFT level, using the PBE[38] exchange-correlation functional and a triple-$\zeta$ plus polarization (TZP) basis set for all atoms. Relativistic effects were included in calculations through the scalar ZORA approximation.[39] At this level of theory, the geometry of organic molecules, gold clusters and and gold-molecule interactions are in excellent agreement with the available data (see **Table S1** in the Supporting Information). DFT calculations were performed with the ADF program package.[40] Electronic structure calculations on mesh grid points around the molecule were performed at the unrestricted PM6 level, using the G09 program package.[41] The binding energy (BE) between gold atoms and clusters and molecules was computed as:

$$BE = E[Au_n@mol] - (E[Au_n] + E[mol])$$

where $E[Au_n@mol]$, $E[Au_n]$ and $E[mol]$ are the energies of the adduct (gold cluster interacting with the molecule), of the gold cluster and of the molecule, respectively. The basis set superposition error (BSSE) was found to be negligible (below 0.03 eV) for the systems considered.

The spatial distribution of the interaction energy between gold atoms and molecules was evaluated by defining a cubic mesh of grid points around the molecule, with a spacing of 1 Å, taking all points within a minimum of 1 Å and a maximum of 4 Å from atoms of the molecule (see **Figure S4** in the Supporting Information). A single organic molecule was considered, as representative of exposed functional moieties in aggregates of molecular materials.

Dynamical effects and the role of thermal processing on the growth of gold cluster at the interface with organic molecules were simulated by applying a method derived from kinetic Monte Carlo techniques. To this end, gold atoms were progressively added to the system, and the interaction energy was evaluated on a cubic mesh of grid points around the molecule, as described previously. At each iterative step, the position of the incoming gold atom was selected according to a Boltzmann distribution (see Supporting Information), on the basis of the total electronic energy of the configuration, and at a temperature of 300K and 1000K, for the simulation of thermodynamically-controlled and kinetically-controlled growth, respectively. The procedure was iterated for the first five gold atoms. The whole set of simulations was repeated ten times with different random number sequences, and the overall statistical distribution of gold atoms was computed. Further details are provided in the Supporting Information section.

**Acknowledgements**
We acknowledge the CINECA award under the ISCRA initiative, for the availability of high-performance computing resources and support.

# Supporting Information

**Early steps in the formation of the interface between organic molecular semiconductors and metals: a computational approach**

*Francesco Mercuri\**

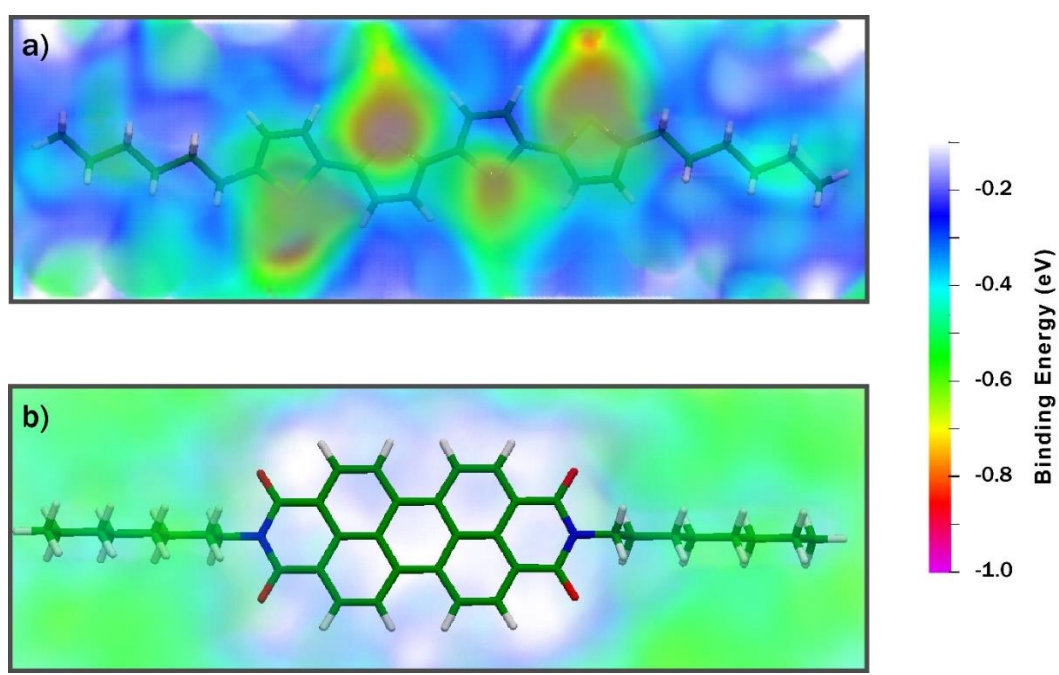

**Figure S1.** Volume rendering of the gold-molecule interaction energies (see scale bar) computed at the PM6 level for a) DH4T and b) PTCDI-C8.

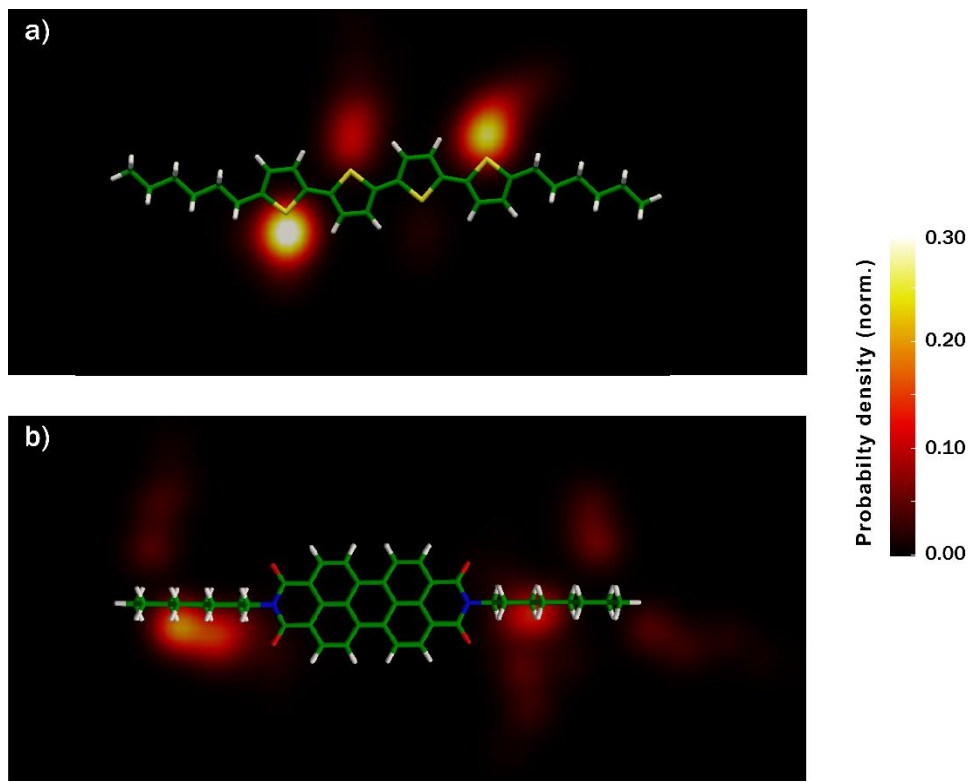

**Figure S2.** Projection on a plane parallel to the molecular π-core of the distribution of positions of gold atoms in the formation of the interface with a) DH4T and b) PTCDI-C8, averaged over the whole ensemble of simulations, in thermodynamic conditions.

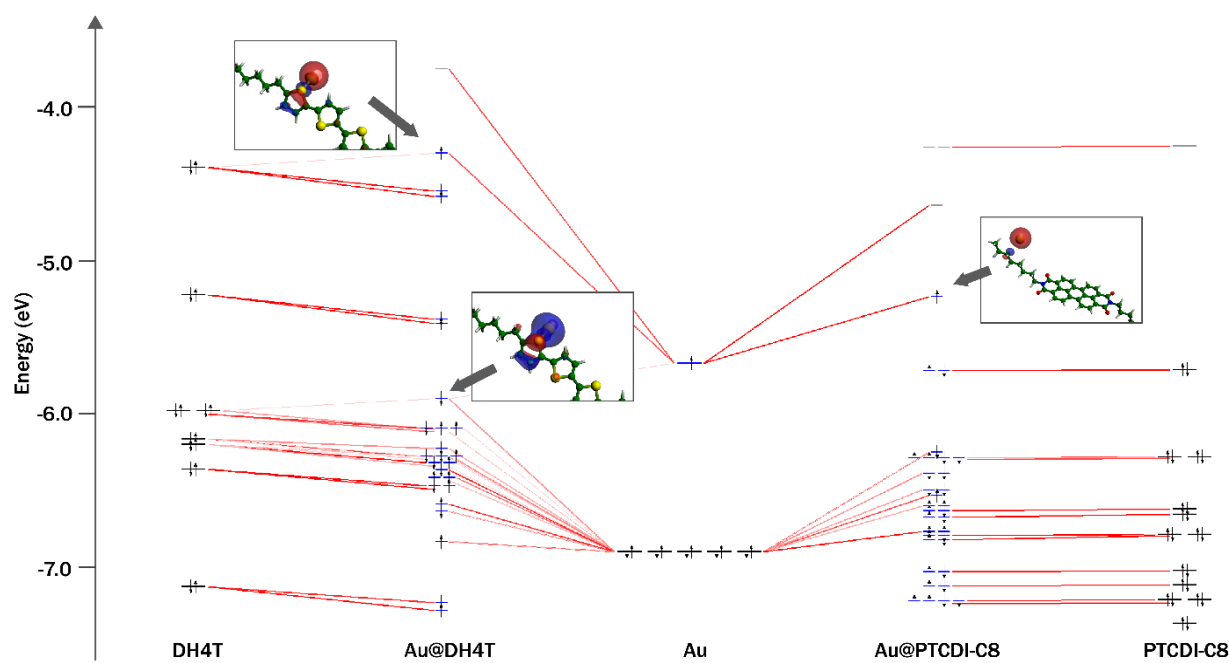

**Figure S3.** Energy level diagram for a single gold atom interacting with DH4T (left) and PTCDI-C8 (right) in the minimum energy configuration. The topology of selected molecular orbitals is also shown.

**Computational details**

***Definition of the mesh grid points for the evaluation of the interaction energies***

The spatial distribution of the interaction energy between gold atoms and an organic molecule was computed by defining a cubic mesh of points around the molecule, as shown in **Figure S4.**

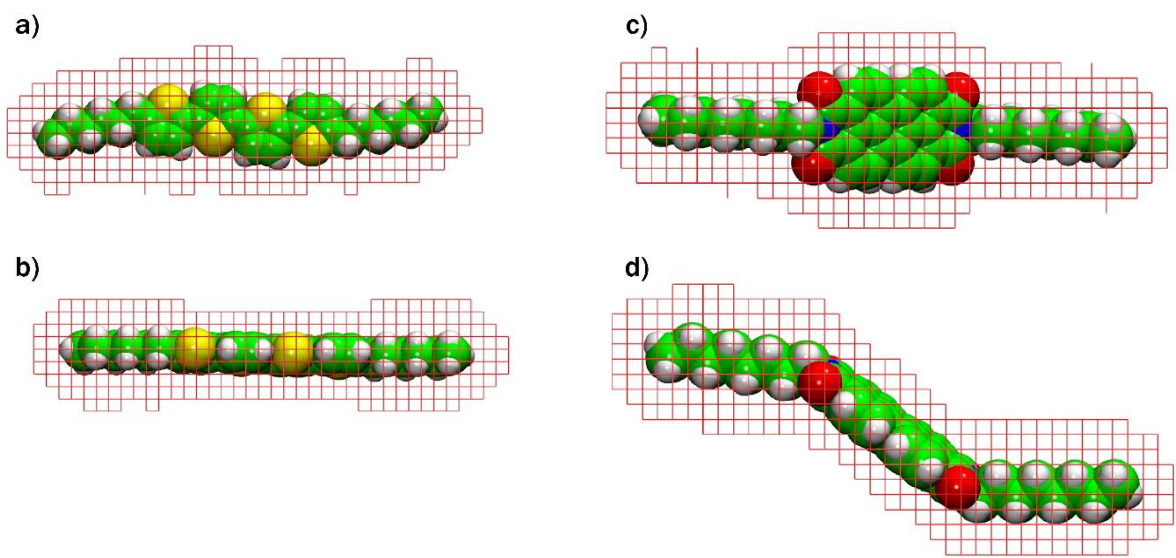

**Figure S4.** Representation of the mesh grid points for the calculation of interaction energies, built around the DH4T (a: top view; b: side view) and PTCDI-C8 (c: top view; d: side view) molecules, respectively.

The mesh grid spacing was set to 1 Å. All points within a minimum distance of 1 Å and a maximum distance of 4 Å from atoms of the molecule were considered. The gold-molecule interaction energy was initially computed at each grid point at both the PM6 and at the DFT level. DFT interaction energies were computed for the initial configuration only (Au@molecule), with the purpose of assessing the accuracy of PM6 methods in evaluating the topology of the spatial distribution of interaction energies around the molecule. In all subsequent iteration steps, PM6 interaction energies, computed at all grid points considered, were used to select the minimum energy configuration. After each iteration step, the position of the gold atom was optimized by constrained-DFT optimization, with thresholds of $10^{-3}$ Ha and $2 \cdot 10^{-3}$ a.u./Å for energy and gradient convergence, respectively. The minimum energy configuration was subsequently used as a starting point for evaluating the interaction with a further gold atom. The interaction energies on grid points were recomputed, at the PM6 level, after each iteration step.

***Simulation of the growth of gold clusters at the interface with molecules in thermodynamically- and kinetically-controlled conditions***

Kinetic effects in early steps in the growth of gold clusters at the interface with organic molecules were simulated by applying a method derived from kinetic Monte Carlo techniques. The energy for a single gold atom interacting with the system was evaluated on a set of mesh grid points around the molecule, as described above. The initial position of the gold atom was therefore selected randomly from a Boltzmann distribution, defined as:

$$p(i) = \frac{e^{-E_i/k_B T}}{\sum_i^N e^{-E_i/k_B T}}$$

Where $p(i)$ is the probability of the configuration $i$ corresponding to a gold atom on a given point of the mesh grid; $E_i$ is the total electronic energy of the configuration; $N$ is the total number of grid points; $k_B$ is the Boltzmann constant and $T$ is the temperature. The temperature of the system was set to 300K and 1000K for simulations of thermodynamically-controlled and kinetically-controlled growth, respectively.

By computing the cumulative function:

$$p_k = \sum_i^k p(i)$$

and the total probability $Q = p_N$, after extracting a uniform random number $u = (0,1]$, the configuration $k$ for which:

$$p_{k-1} < uQ \leq p_k$$

was selected.[6] The position of the gold atom in the selected configuration was optimized by constrained DFT after each iteration step, as described above, and the interaction energy of a new incoming gold atom on the grid points was recomputed at the PM6 level. The procedure was repeated iteratively for the first five gold atoms added sequentially to the isolated organic molecule. The whole set of iterations was repeated ten times for each molecule (DH4T and PTCDI-C8, respectively) and for each of the growth conditions (thermodynamically- and kinetically-controlled, respectively), changing each time the seed of the random number generator. For each molecule in a given growth condition, the positions of the gold atoms around the molecule (a total of 50 atoms for each set of simulations, corresponding to the sequence of the first five gold atoms added consecutively to the molecule, repeated ten times with different random number sequences) were collected. The distribution of likely interaction sites around the molecule at a given temperature in the early steps of the formation of the gold-molecule interface, was smoothed by Gaussian filtering (with a width of $\sigma=1$ Å) on a three-dimensional grid mesh of the collected set of atom positions. The relevance of initial steps in the formation of the metal-molecule interface was reproduced by assigning a larger weight to the first interacting atoms. The 3D grid of smoothed atom distributions was projected onto a plane by integration along a direction orthogonal to the $\pi$ core of the molecule. The resulting 2D plot represents the importance-weighted mean distribution of gold atoms around the molecule, projected onto a plane that is parallel to the molecular $\pi$ core.

**Table S1.** Comparison of relevant geometrical parameters with available experimental and computed data. Distances are given in Å, angles in degrees.

| | | This work | Exp. |
|---|---|---|---|
| DH4T | $C_\alpha$-S | 1.75 | 1.73[a] |
| | $C_\alpha$-$C_\beta$ | 1.39 | 1.37 |
| | <C-S-C | 92.3 | 92.2 |
| PTCDI-C8 | C-N | 1.41 | 1.40[b] |
| | C-O | 1.23 | 1.24 |
| | $C_{imide}$-$C_3$ | 1.48 | 1.48 |
| Au@DH4T | Au-S | 2.48 | 2.5-2.75[c] |
| $Au_{10}$@DH4T | Au-Au | 2.78 | 2.72[d] |
| $Au_{10}$@PTCDI-C8 | Au-Au | 2.75 | 2.72[d] |

[a]Values for quaterthiophene, taken from Ref.[14]

[b]Taken from Ref. [15]

[c]Taken from Ref. [16]

[d]Values computed for the $Au_{10}$ cluster at the PBE/def2–TZVP level, taken from Ref.[34]